\documentclass[prl,aps,twocolumn,amsmath,showpacs,amssymb]{revtex4}
\usepackage{graphicx}% Include figure files
\usepackage{dcolumn}% Align table columns on decimal point
\usepackage{bm}% bold math

\begin{document}
\title{Anisotropic Magnetoresistance in Charge-Ordering $Na_{0.34}(H_3O)_{0.15}CoO_2$:
Strong Spin-Charge Coupling and Spin Ordering}

\author{C. H. Wang$^1$}
\author{ X. H. Chen$^1$}
\altaffiliation{Corresponding author} \email{chenxh@ustc.edu.cn}
\author{G. Wu$^1$, T. Wu$^1$, H. T. Zhang$^1$, J. L. Luo$^2$, G. T. Liu$^2$ and N. L. Wang$^2$}
\affiliation{ 1. Hefei National Laboratory for Physical Science at
Microscale and Department of Physics, University of Science and
Technology of China, Hefei, Anhui 230026, People's Republic of
China} \affiliation{2. Beijing National Laboratory for Condensed
Matter Physics, Institute of Physics, Chinese Academy of Science,
Beijing 100080, People's Republic of China}

\date{\today}

\begin{abstract}
Angular-dependent in-plane magnetoresistance (AMR) for single
crystal $Na_{0.34}(H_3O)_{0.15}CoO_2$ with charge ordering is
studied systematically. The anisotropic magnetoresistance shows a
twofold symmetry at high temperature with rotating H in the Co-O
plane, while a sixfold symmetry below a certain temperature
($T_\rho$). At $T_\rho$, the symmetry of AMR changes from twofold
to fourfold with rotating magnetic field (H) in the plane
consisting of the current and c-axis. The variation of AMR
symmetry with temperature arises from the subtle changes of the
spin structure. These results give a direct evidence for the
itinerant electrons directly coupled to the localized spins.
\end{abstract}


\maketitle
\newpage

The layered sodium cobaltate $Na_xCoO_2$ has become one of the
focus in research due to the discovery of superconductivity with
$T_c\sim 5$ K in $Na_{0.35}CoO_2\cdot1.3H_2O$.\cite{Takada} In
fact, the phase diagram in this system is very
rich\cite{foo,Sakurai,Milne} and one has observed  charge order
and several complicated magnetic orders in addition to
superconductivity.\cite{Sugiyama,Luojl,Mendels}  Recently, many
anomalous properties were observed in this system.  The
angle-resolved photoemission spectroscopy(ARPES) experiments
indicates that in $Na_{0.7}CoO_2$ the well-defined quasiparticle
peaks only exist in low temperature when a coherent transport
forms along c-axis.\cite{hasan} Such peculiar behaviors make one
believe that $Na_xCoO_2$ could be regarded as a new prototype for
studying the doped Mott insulator.

$Na_xCoO_2$ system shows complicated magnetic structure. The bulk
antiferromagnetism (AF) with the ordering moment perpendicular to
Co-O plane was observed in $Na_{0.82}CoO_2$\cite{Bayrakci}. In
$Na_{0.75}CoO_2$, the ferromagnetic correlation within the cobalt
layer and AF correlation between the cobalt layers were observed
with spin direction along c-axis.\cite{Boothroyd,Helme}  Neutron
diffraction and NMR studies indicate that there exist two distinct
Co sites with different magnetic moments in $Na_{0.5}CoO_2$ and
antiferromagnetic ordering occurs below $T_{c1}\sim 87$
K.\cite{yokoi,gasparovic} However, two different magnetic
structures are proposed by Yokoi et al.\cite{yokoi} and Gasparovic
et al.\cite{gasparovic}, respectively. Gasparovic et al. proposed
that magnetic moment of $Co^{4+}$ ions that is magnetic (in the
S=1/2 low spin state) aligns antiferromagnetically, the rows of
$Co^{4+}$ ions alternate with rows of $Co^{3+}$ ions (in the S=0
spin state) that are nonmagnetic. Yokoi et al. believed that the
spin structure of $Na_{0.5}CoO_2$ has three axis, two axes are
$Co^{3.5-\delta}-Co^{3.5-\delta}$ and the other one is
$Co^{3.5+\delta}-Co^{3.5+\delta}$. The large moments of
$Co^{3.5+\delta}$ sites align antiferromagnetically with spin
direction in Co-O plane. The small moments of $Co^{3.5-\delta}$
sites is along c-axis and their interplane correlation is
antiferromangetic. The magnetic structure for $Na_{0.5}CoO_2$ is
open question.

The oxonium ions, $(H_3O)^+$, can occupy the same crystallographic
sites of the Na ions when the $Na_xCoO_2$ crystal is immersed in
distilled water.\cite{takada1,goodenough} Intercalation of
$(H_3O)^+$ makes the Co valence to be lower than theoretical value
estimated directly from Na content. Much research focused on the
effect of oxonium ions\cite{goodenough,Milne} in $Na_xCoO_2$
system, so that superconducting phase diagram is revised.
Oxidation state of Co for superconducting sample is very close to
3.5 of $Na_{0.5}CoO_2$ due to intercalation of oxonium ions
instead of about 3.7.\cite{karppinen} One should take the charge
ordering insulator with Co oxidation state of $\sim3.5$ as parent
compound for the superconductor. Therefore, understanding on
microscopic magnetic structure of $Na_{0.5}CoO_2$ is very
important to understand the charge ordering behavior and
superconducting mechanism.

Magnetoresistance (MR) provides new insight into the coupling
between charges and background magnetism. This is particularly
valuable because, as shown in this work, small magnetic moments
order in the background of magnetic ordering with large magnetic
moments, and such ordering is difficultly detected by
magnetization measurement. In this paper, we studied AMR of
$Na_{0.34}(H_3O)_{0.15}CoO_2$ crystal with the same charge
ordering as that in $Na_{0.5}CoO_2$.\cite{chenxh} Change of AMR
symmetry from twofold to sixfold is observed with rotating H in
the Co-O plane at a certain temperature ($T_\rho$). At $T_\rho$,
symmetry of AMR changes from twofold to fourfold with rotating
magnetic field (H) in the plane consisting of the current (I) and
c-axis. Such symmetry change of AMR can be well understood by spin
ordering of $Co^{3.5-\delta}$ sites with small magnetic moments
below $T_\rho$. Our results support the model proposed by Yokoi et
al.\cite{yokoi} These results give a direct evidence for the
itinerant electrons directly coupled to the localized spins.

High quality single crystal $Na_{0.34}(H_3O)_{0.15}CoO_2$ was
obtained through $Na_{0.41}CoO_2$ crystal (20 mg) immersed in
distilled water (120 ml) for about 120 hours. The actual Na
concentration was determined by inductively coupled plasma
spectrometer (ICP) chemical analysis. The detailed description on
the synthesis  and characterization of
$Na_{0.34}(H_3O)_{0.15}CoO_2$, which is only sample to show the
charge ordering behavior except for $Na_{0.5}CoO_2$, has been
reported elsewhere.\cite{chenxh} The near-normal incident
reflectance spectra were measured on the freshly cleaved surface
by a Bruker 66 v/s spectrometer in the frequency range from
40$\sim$2900 $cm^{-1}$, as described in our early
report.\cite{wangnl} Standard Kramers-Kronig transformations were
employed to drive the frequency-dependent conductivity spectra.
Susceptibility and magnetoresistance were measured in Quantum
Design SQUID and PPMS system, respectively.

Fig. 1(a) shows temperature dependence of in-plane resistivity and
susceptibility for $Na_{0.34}(H_3O)_{0.15}CoO_2$. As shown in
Fig.1(a), similar insulating behavior to that in $Na_{0.5}CoO_2$
was observed below 50 K. It indicates that charge ordering
behavior occurs in $Na_{0.34}(H_3O)_{0.15}CoO_2$. Susceptibility
shows two transitions at 84 and 47 K. It is quite similar to that
observed in $Na_{0.5}CoO_2$ although the transition temperatures
are slightly lower compared to $Na_{0.5}CoO_2$. Oxidation state
(3.49) of Co ions in $Na_{0.34}(H_3O)_{0.15}CoO_2$ is almost the
same as that (3.50) in $Na_{0.5}CoO_2$.  It suggests that the
charge ordering in $Na_xCoO_2$ system strongly depends on the
valence of Co ion. In comparison with $Na_{0.5}CoO_2$,
differential curve $d\rho/dT$ was shown in Fig.1(b) and (c). A dip
of the $d\rho/dT$ was observed at $T_{c1}\sim 84$ K, and the
obvious change in the slope $d\rho/dT$ was also observed at
$T_{c2} \sim 50 $ K and $T_{\rho} \sim 10$ K, respectively. These
features are quite similar to that in $Na_{0.5}CoO_2$, in which
the dip is observed at $T_{c1} \sim 87$ K and the slope of
$\rho(T)$ increases abruptly at $T_{c2} \sim 53$ K and $T_{\rho}
\sim 20 $ K.\cite{foo,huang,wangch} It is commonly believed that
the anomaly at $T_{c1} \sim 87$ K arises from an antiferromagnetic
ordering.\cite{yokoi,gasparovic} The anomaly at $T_{c2} \sim 53$ K
is not clear although a kink is observed in
susceptibility\cite{foo,huang} due to no change in magnetic
structure drastically.\cite{yokoi} Dramatic change in $d\rho/dT$
at $T_{\rho}$ is believed to arise from spin
reorientation.\cite{huang} The behavior in
$Na_{0.34}(H_3O)_{0.15}CoO_2$ crystal is similar to
$Na_{0.5}CoO_2$ despite these anomalous temperatures are lower.
Therefore, the origin for these anomalies in
$Na_{0.34}(H_3O)_{0.15}CoO_2$ and $Na_{0.5}CoO_2$ should be the
same.

\begin{figure}[h]
\includegraphics[width=8cm]{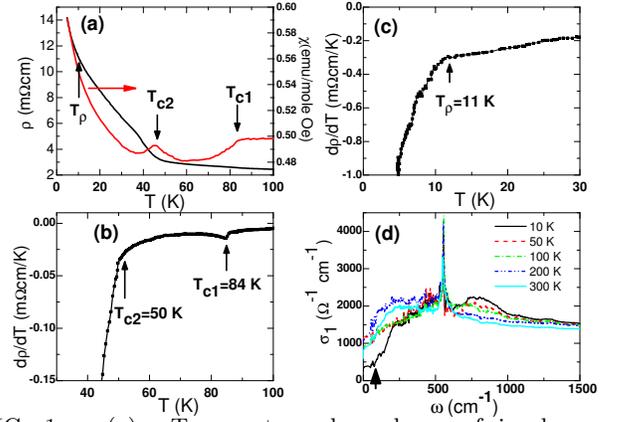}\vspace{-5mm}
\caption{\label{fig:epsart} (a): Temperature dependence of
in-plane resistivity and susceptibility with H $\|$ Co-O plane for
$Na_{0.34}(H_3O)_{0.15}CoO_2$; Temperature dependence of the slope
$d\rho/dT$ for the data shown in (a), (b): temperature range from
30 to 100 K; (c): the temperatures below 30 K; (d): In-plane
optical conductivity spectra of $Na_{0.34}(H_3O)_{0.15}CoO_2$.}
\end{figure}

To further characterize the charge ordering behavior in
$Na_{0.34}(H_3O)_{0.15}CoO_2$, in-plane optical conductivity
spectra are measured. As shown in Fig.1(d), in-plane optical
conductivity spectra of $Na_{0.34}(H_3O)_{0.15}CoO_2$ are quite
similar to that of $Na_{0.5}CoO_2$.\cite{wangnl} A broad hump
appears below 100 K at about 800 $cm^{-1}$ was
observed.\cite{wangnl} A sharp suppression of the conductivity
spectra below charge ordering temperature indicates an opening of
charge gap. The gap value is close to the reported value of $\sim
125$ $cm^{-1}$ with $2\Delta/k_BT_{co}=3.5$.\cite{wangnl} It
further indicates that all features observed in $Na_{0.5}CoO_2$
occur in $Na_{0.34}(H_3O)_{0.15}CoO_2$.

Fig. 2(a) shows isothermal in-plane MR at 5 K and 20 K with
different $\alpha$. The $\alpha$ is the angle between magnetic
field H and current I.  It can be seen that MR are negative at 5K
and 20 K with $H \| I$ (within Co-O plane), while positive with $H
\bot$ the Co-O plane.  One should note that MR is almost zero with
$\alpha=90^o$ at 20 K. Such behavior has been observed in
$Na_{0.5}CoO_2$.\cite{wangch}

\begin{figure}[h]
\includegraphics[width=8cm]{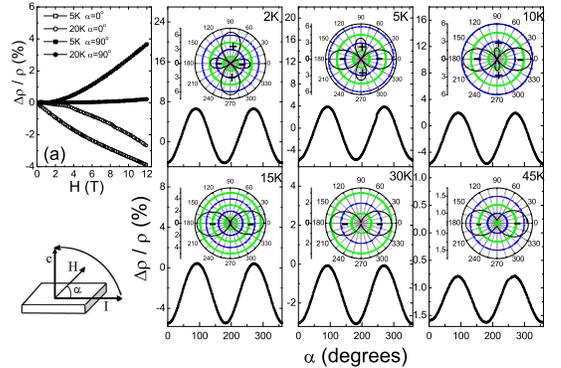}\vspace{-5mm}
\caption{\label{fig:epsart} (a): The isothermal MR at 5 K and 20 K
with different angles between H and I; (b): The angular dependent
isothermal MR and polar plot for $Na_{0.34}(H_3O)_{0.15}CoO_2$
under H=12 Tesla at different temperatures. The magnetic field H
is kept in the plane consisting of the current and c-axis when the
sample is rotated, as shown in schematic diagram.}
\end{figure}

In order to study effect of spin ordering on the charge transport
(especially below $T_{\rho}$) and to understand the behavior
observed in Fig.2(a), we investigated the angular dependent
in-plane MR behavior at different temperatures by rotating H in
the plane consisting of the current and c-axis, as shown in
schematic diagram. The magnetic field is fixed at 12 T during the
measurement. As shown in Fig.2, the MR increases with decreasing
temperature, and the magnitude of MR $\Delta\rho_{ab}/\rho_{ab}$
at 2 K is almost five times larger than that at 45 K. The data
reveal that the AMR has two-fold symmetry above 15 K, and the MR
is always negative and the maximum negative MR occurs at
$\alpha$=0 and 180$^o$, that is: the H lies in the Co-O plane.
While the MR is almost zero with H perpendicular to the Co-O plane
above 15 K. A striking feature is that the AMR shows clearly a
symmetry evolution from p-wave-like to d-wave-like with decreasing
temperature. One can observe only two negative arms above 15 K.
Two small positive arms begin to be observed at 10 K. The two
positive arms become large with further decreasing temperature,
and eventually dominate at 2 K. With the growth of positive arms,
one can find that the negative arms shrink slightly. It should be
pointed out that the AMR symmetry transition from twofold to
fourfold occurs at $\sim 15$ K close to $T_{\rho}$. Such
fascinating transition was also observed at $T_{\rho}\sim 20$ K in
$Na_{0.5}CoO_2$.\cite{wangch} Such d-wave like behavior in AMR has
been observed in antiferromagnetic $YBa_2Cu_3O_{6+x}$, and is
believed as a consequence of the rotation of stripe direction with
respect to the current direction.\cite{ando1} Therefore, we
believe that the evolution of AMR symmetry observed in charge
ordering state should arise from change of magnetic structure
under magnetic field.

 AMR is also studied with rotating H in the Co-O plane.
Fig.3 shows isothermal in-plane MR as a function of angle and the
polar plot for $Na_{0.34}(H_3O)_{0.15}CoO_2$ at H=12 Tesla at
different temperatures. The magnitude of MR increases
approximately by 5 times at low temperature relative to that at 40
K. Similar to the case with rotating H in the plane consisting of
the current and c-axis, symmetry of AMR is twofold above 15 K, but
changes to sixfold at 2 K. The variation of MR with angle follows
a sine wave above 15 K. The data at 15 K clearly show that peak
and valley of the sine wave broaden and become asymmetry. It
suggests that symmetry of AMR begins to change to sixfold at 15 K,
which is consistent with the temperature corresponding to twofold
to fourfold symmetry in AMR as shown in Fig.2. It implies that
symmetry change of AMR in Fig.2 and Fig.3 arises from the same
origin. Nevertheless, the data in Fig.3 exhibit some distinct
difference from that in Fig. 2. Firstly, the MR remains negative
at all angles in the whole temperature regime, and the magnitude
of MR at each fixed angle always increases with decreasing
temperature. Secondly, AMR shown in Fig.3 changes to six-fold at 2
K and exhibits obviously asymmetric, while AMR at various
temperatures shown in Fig.2 is symmetric.

\begin{figure}[h]
\includegraphics[width=8cm]{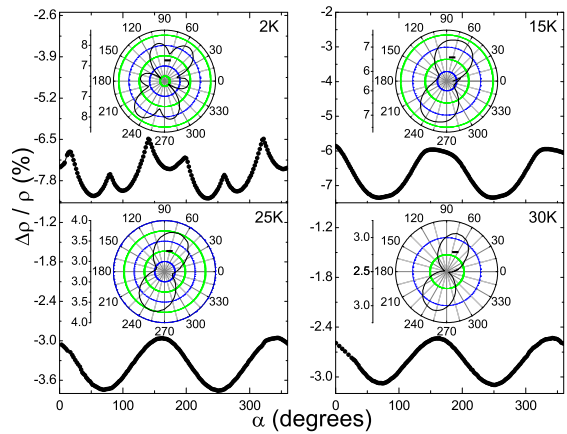}
\caption{\label{fig:epsart} The isothermal MR as a function of
angle for the $Na_{0.34}(H_3O)_{0.15}CoO_2$ at H=12 Tesla and the
polar plot. H is kept in Co-O plane when the sample is rotated.}
\end{figure}

Another striking feature is that the angle difference between
neighboring peaks in AMR at 2 K shown in Fig.3 is exactly 60$^o$.
Therefore, such nice sixfold symmetry of AMR at 2 K seems to be
related to the crystallographic structure because the crystal
structure of $Na_xCoO_2$ is hexagonal. However, the sixfold
symmetry of AMR with rotating H in Co-O plane is only observed
below 15 K as shown in Fig.3. Therefore, such sixfold symmetry of
AMR does not originate directly from the crystal structure. It
should be pointed out that no such sixfold symmetry of AMR can be
observed in other $Na_xCoO_2$ without charge ordering. It suggests
that the symmetric feature of AMR observed in Fig.2 and Fig.3
could be related to the spin structure. Yokoi et al. proposed a
magnetic structure shown in Fig.4 (a) and (b) for $Na_{0.5}CoO_2$.
The spin structure has three axis, two axes are
$Co^{3.5-\delta}-Co^{3.5-\delta}$ and the other one is
$Co^{3.5+\delta}-Co^{3.5+\delta}$.\cite{yokoi} The large moments
of $Co^{3.5+\delta}$ sites align antiferromagnetically at
$T_{c1}\sim 87$ K with spin direction in Co-O plane. The small
moments of $Co^{3.5-\delta}$ sites is along c-axis and their
interplane correlation is antiferromangetic. It is difficult to
distinguish their in-plane correlation, so that there exist two
possibilities for the in-plane correlation as shown in Fig.4(a)
and (b).\cite{yokoi} But they remain an open question that at what
temperature the small moments of $Co^{3.5-\delta}$ sites order.
The evolution of the AMR symmetry with temperature shown in Fig.2
and Fig.3 can be well understood by the spin structure proposed by
Yokoi et al. At high temperature, the symmetry of AMR is twofold
for both of the cases with rotating H in the Co-O plane and in the
plane consisting of the c-axis and the current. This is because
only the large moments of $Co^{3.5+\delta}$ sites order
\begin{figure}[h]
\includegraphics[width=8cm]{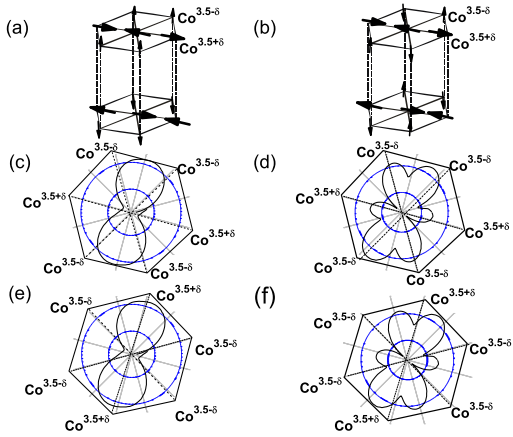}
\caption{\label{fig:epsart} (a) and (b): Possible magnetic
structures proposed by Yokoi et al.\cite{yokoi} for
$Na_{0.5}CoO_2$; (c) and (d): A possible configuration for the
in-plane matching between the magnetic structure and the AMR
symmetry  at 15 and 2 K, respectively; (e) and (f): Another
possible configuration for the in-plane matching between the
magnetic structure and the AMR symmetry  at 15 and 2 K,
respectively.}
\end{figure}
antiferromagnetically at high temperature. Therefore, only a
"stripe"-like spin ordering along
$Co^{3.5+\delta}-Co^{3.5+\delta}$ exists in the Co-O plane. It
naturally explains the twofold symmetry of AMR shown in Fig.2 and
Fig.3. For the case with rotating H in Co-O plane at high
temperatures, we proposed a model for the matching between the AMR
symmetry and the magnetic structure shown in Fig.4(c) and (e).
There exist two possibilities for the relation between the AMR
symmetry and magnetic structure: one is that the maximum MR occurs
with H along the $Co^{3.5+\delta}-Co^{3.5+\delta}$ axis, the other
is perpendicular to the $Co^{3.5+\delta}-Co^{3.5+\delta}$ axis. At
low temperatures, the spin ordering at $Co^{3.5-\delta}$ sites
with small moment occurs, and consequently the magnetic structure
is hexagonal in the Co-O plane. Such hexagonal magnetic structure
leads to the sixfold symmetry of AMR with rotating H in Co-O plane
shown in Fig.3. The two possible correlations between the AMR
symmetry and the magnetic structure are proposed in Fig.4(d) and
(f) at low temperatures. Based on the observation of sixfold
symmetry of AMR at 2 K shown in Fig.3, the magnetic structure
prefers to that in-plane correlation for the magnetic moments at
$Co^{3.5-\delta}$ sites is antiferromagnetic shown in Fig.4(b).
This is because in-plane symmetry of the magnetic structure is not
hexagonal if all the $Co^{3.5-\delta}$ sites are exactly
equivalent as shown in Fig.4(a). The fourfold and twofold
symmetries of AMR shown in Fig.2 can be also understood with the
magnetic structures at low and high temperatures, respectively.
Therefore, present results support the magnetic structure proposed
by Yokoi et al.\cite{yokoi} The magnetic structure proposed by
Gasparovic et al.\cite{gasparovic} cannot explain the sixfold
symmetry of AMR observed in Fig.3 because only a "stripe"-like
spin ordering along $Co^{4+}-Co^{4+}$ occurs in Co-O plane in
their model. A strong scattering of electrons from spins should
occur with the current perpendicular to the antiferromagnetic
"stripe" along $Co^{3.5+\delta}-Co^{3.5+\delta}$. However, it is
difficult to distinguish which set of the configurations shown in
Fig.4(c) and (d), Fig.4(e) and (f) is more realistic because we
cannot determine the $Co^{3.5+\delta}-Co^{3.5+\delta}$ axis. The
evolution of the AMR symmetry with temperature shown in Fig.2 and
Fig.3 suggests that the spin ordering of $Co^{3.5-\delta}$ sites
with small moments occurs at $\sim 15$ K, which is close to the
$T_{\rho}$. This is explained why the resistivity increases
rapidly at $T_{\rho}$. It should be pointed out that symmetry
change of AMR observed in $Na_{0.34}(H_3O)_{0.15}CoO_2$ can be
also observed in $Na_{0.5}CoO_2$,  only $T_{\rho}$ $\sim 20$ K at
which the ARM symmetry changes from twofold to sixfold is slightly
different in $Na_{x}CoO_2$ x$\sim$ 0.5.\cite{hu}. These results
give a good understanding on the transport and magnetic properties
for the charge ordering $Na_xCoO_2$, especially on the anomalies
at $T_{\rho}$, $T_{c1}$ and $T_{c2}$.

In this letter, the angular-dependent in-plane magnetoresistance
(AMR) for $Na_{0.34}(H_3O)_{0.15}CoO_2$ is systematically studied.
The symmetry of AMR is twofold for both of the case with rotating
H in the Co-O plane and in the plane consisting of the current and
c-axis above $T_{\rho}$, while changes to sixfold and fourfold
below $T_{\rho}$, respectively. These intriguing results support
the magnetic structure proposed by Yokoi et al. We solve the open
question for the spin ordering temperature ($\sim$ $T_{\rho}$) of
the small moments at $Co^{3.5-\delta}$ sites. The present work
shows that the AMR is an auxiliary but powerful tool to study the
subtle change of the magnetic structure. The magnetic and
transport properties of $Na_{0.34}(H_3O)_{0.15}CoO_2$ are the same
as those of $Na_{0.5}CoO_2$. It indicates intercalation of oxonium
ions occurred and its charge compensation to make the oxidation
state of Co to be close to $\sim3.5$ when $Na_xCoO_2$ crystal was
hydrated. Therefore, one should take the charge ordering insulator
with Co oxidation state of $\sim3.5$ as parent compound of the
superconductor to think about superconducting mechanism and phase
diagram.

 {\bf Acknowledgement:} This work is
supported by the Nature Science Foundation of China and by the
Ministry of Science and Technology of China (973 project No:
2006CB601001), and by the Knowledge Innovation Project of Chinese
Academy of Sciences.


\begin{thebibliography}{00}

\bibitem{Takada} K. Takada et al.,  Nature \textbf{422}, 53(2003).
\bibitem{foo} M. L. Foo et al., Phys. Rev. Lett. \textbf{92}, 247001(2004).
\bibitem{Sakurai} H. Sakurai et al., J. Phys. Soc. Jpn. \textbf{74}, 2909(2005).
\bibitem{Milne} C. J. Milne et al., Phys. Rev. Lett. \textbf{93},
247007(2004).
\bibitem{Sugiyama} J. Sugiyama et al., Phys. Rev. Lett. \textbf{92}, 017602(2004).
\bibitem{Luojl} J. L. Luo et al., Phys. Rev. Lett. \textbf{93}, 187203(2004).
\bibitem{Mendels} P. Mendels et al., Phys. Rev. Lett. \textbf{94}, 136403(2005).
\bibitem{hasan}  M. Z. Hasan et al., Phys. Rev. Lett. \textbf{92}, 246402(2004)
\bibitem{Bayrakci} S. P. Bayrakci et al.,  Phys. Rev. B \textbf{69}, 100410(2005)
\bibitem{Boothroyd} A. T. Boothroyd et al.,  Phys. Rev. Lett. \textbf{92},
197201(2004).
\bibitem{Helme}
L. M. Helme et al., Phys. Rev. Lett. \textbf{94}, 157206(2005).
\bibitem{yokoi} M. Yokoi et al., J. Phys. Soc. Jpn {\bf 74},
3046(2005).
\bibitem{gasparovic}G. Gasparovic et al., Phys. Rev. Lett. {\bf 96}, 046403(2006).
\bibitem{takada1} K. Takada et al., J. Mater. Chem. \textbf{14}, 1448(2004).
\bibitem{goodenough} M. Banobre-Lopez et al., Chem. Mater. \textbf{17},
1965(2005).
\bibitem{karppinen} M. Karppinen et al., Chem. Mater. {\bf 16},
1693(2004); P. W. Barnes et al., Phys. Rev. B {\bf 72},
134515(2005).
\bibitem{chenxh}  C. H. Wang et al., Solid State Commun. {\bf 138}, 169(2006).
\bibitem{huang} Q. Huang et al., J. Phys.: Condens. Matter \textbf{16}, 5803(2004)
\bibitem{wangch} C. H. Wang et al., Phys. Rev. B \textbf{71}, 224515(2005).
\bibitem{wangnl} N. L. Wang et al., Phys. Rev. Lett. \textbf{93}, 147403(2004).
\bibitem{ando1}
Y. Ando, A. N. Lavrov, and S. Komiya,   Phys. Rev. Lett. {\bf 83},
2813(2003).
\bibitem{hu}
F. Hu et al., Phys. Rev. B {\bf 73}, 212414(2006).









\end{thebibliography}
\end{document}